\documentclass[numberedappendix,12pt,preprint]{emulateapj}

\usepackage[]{amsmath}

\shorttitle{Through The Looking GLASS}
\shortauthors{Schmidt et al. (2014)}

\usepackage{color}			      
\definecolor{midgray}{gray}{0.4}		
\definecolor{orange}{rgb}{1,0.5,0}    

\usepackage[colorlinks=true,citecolor=midgray,linkcolor=midgray]{hyperref}        

\newcommand{\simgt}{\,\rlap{\lower 3.5 pt \hbox{$\mathchar \sim$}} \raise
1pt \hbox {$>$}\,}
\newcommand{\simlt}{\,\rlap{\lower 3.5 pt \hbox{$\mathchar \sim$}} \raise
1pt \hbox {$<$}\,}

\newcommand{\Om}{\Omega_{\rm m}}
\newcommand{\OL}{\Omega_{\Lambda}}

\newcommand{\lya}{Ly$\alpha$}

\newcommand{\tnm}[1]{$^\textrm{#1}$}
\newcommand{\Nhigh}{21}

\newcommand{\Nspec}{14}

\newcommand{\Nok}{9}
\newcommand{\Nmul}{4}

\newcommand{\HST}{\textit{HST}}

\newcommand{\M}{MACSJ0717.5+3745}
\newcommand{\cgs}{erg s$^{-1}$cm$^{-2}$}

\begin{document}


\title{Through the looking GLASS: \HST\ spectroscopy of faint galaxies lensed by the Frontier Fields cluster \M}


\author{
K.~B.~Schmidt$^{1}$, T.~Treu$^{1}$, G.~B.~Brammer$^2$, M.~Brada\v{c}$^3$,
X.~Wang$^{1}$, M.~Dijkstra$^4$, A.~Dressler$^5$, A.~Fontana$^6$,
R.~Gavazzi$^7$, A.~L.~Henry$^8$, A.~Hoag$^3$, T.~A.~Jones$^{1}$,
P.~L.~Kelly$^9$, M.~A.~Malkan$^{10}$, C.~Mason$^{1}$,
L.~Pentericci$^6$, B.~Poggianti$^{11}$, M.Stiavelli$^2$,
M.~Trenti$^{12}$, A.~von der Linden$^{13,14}$, B.~Vulcani$^{15}$}
\affil{$^{1}$ Department of Physics, University of California, Santa Barbara, CA, 93106-9530, USA}
\affil{$^{2}$ Space Telescope Science Institute, 3700 San Martin Drive, Baltimore, MD, 21218, USA}
\affil{$^3$ Department of Physics, University of California, Davis, CA, 95616, USA}
\affil{$^4$ Institute of Theoretical Astrophysics, University of Oslo, Postboks 1029, 0858 Oslo, Norway}
\affil{$^5$ The Observatories of the Carnegie Institution for Science, 813 Santa Barbara St., Pasadena, CA 91101, USA}
\affil{$^6$ INAF - Osservatorio Astronomico di Roma Via Frascati 33 - 00040 Monte Porzio Catone, I}
\affil{$^7$ Institute d'Astrophysique de Paris, F}
\affil{$^8$ Astrophysics Science Division, Goddard Space Flight Center,
Code 665, Greenbelt, MD 20771}
\affil{$^9$ Department of Astronomy, University of California, Berkeley, CA 94720-3411, USA}
\affil{$^{10}$ Department of Physics and Astronomy, UCLA, Los
Angeles, CA, USA 90095-1547}
\affil{$^{11}$ INAF-Astronomical Observatory of Padova, Italy}
\affil{$^{12}$ Institute of Astronomy and Kavli Institute for Cosmology, University of Cambridge, Madingley Road, Cambridge, CB3 0HA, UK} 
\affil{$^{13}$Dark Cosmology Centre, Niels Bohr Institute, University of Copenhagen Juliane Maries Vej 30, 2100 Copenhagen {\O}, DK}
\affil{$^{14}$ Kavli Institute for Particle Astrophysics and Cosmology, Stanford University, 452 Lomita Mall, Stanford, CA  94305-4085, USA}
\affil{$^{15}$ Kavli Institute for the Physics and Mathematics of the Universe (WPI), Todai
Institutes for Advanced Study, the University of Tokyo, Kashiwa, 277-8582,
Japan}
\email{kschmidt@physics.ucsb.edu}

\begin{abstract}
The \emph{Grism Lens-Amplified Survey from Space} (GLASS) is a Hubble
Space Telescope (\HST) Large Program, which will obtain 140 orbits of
grism spectroscopy of the core and infall regions of 10 galaxy
clusters, selected to be among the very best cosmic telescopes.
Extensive \HST\ imaging is available from many sources including the
CLASH and Frontier Fields programs.  We introduce the survey by
analyzing spectra of faint multiply-imaged galaxies and $z\gtrsim6$
galaxy candidates obtained from the first seven orbits out of fourteen targeting the
core of the Frontier Fields cluster \M. 
Using the G102 and G141 grisms to cover the wavelength range 0.8--1.7$\mu$m, 
we confirm \Nmul\ strongly lensed systems by
detecting emission lines in each of the images.  For the
\Nok\ $z\gtrsim6$ galaxy candidates clear from contamination,
we do not detect any emission lines down to a seven-orbit 1$\sigma$ noise level of
$\sim$5$\times$10$^{-18}$\cgs. Taking lensing magnification into
account, our flux sensitivity reaches
$\sim$0.2--5$\times$10$^{-18}$\cgs.  These limits over an
uninterrupted wavelength range rule out the possibility that the
high-$z$ galaxy candidates are instead strong line emitters at lower
redshift.  These results show that by means of careful modeling of the
background --- and with the assistance of lensing
magnification --- interesting flux limits can be reached for large
numbers of objects, avoiding pre-selection and the wavelength
restrictions inherent to ground-based multi-slit spectroscopy. These
observations confirm the power of slitless \HST\ spectroscopy even in
fields as crowded as a cluster core.
\end{abstract}



\section{Introduction}
\label{sec:intro}

The emergence of the first galaxies from the mist of cosmic dawn is
one of the major outstanding questions in current
astrophysics. Measurements of the cosmic microwave background
anisotropy indicate that the universe was reionized during the
redshift range $z\sim$7--12
\citep[e.g.][]{PlanckCollaboration:2013p33616,Bennett:2013p27238,Hinshaw:2013p27234}. The
first galaxies were most likely the source of reionizing photons,
although this has not been conclusively proven given the observational
and theoretical uncertainties
\citep{Robertson:2013p27340,Schmidt:2014p33431}.

Finding and spectroscopically confirming the first galaxies is
essential not only to identify the sources of reionization but also to
understand the physical mechanisms at work in the early starforming
regions, the interstellar and circumgalactic media. Much progress has
been achieved with photometric studies. Large samples of candidate
galaxies at $z\gtrsim8$ have been identified in blank fields or behind
clusters using the Lyman break technique or spectral energy
distribution fitting techniques
\citep{Bouwens:2010p30142,Ellis:2012p26700,Oesch:2013p27877,Coe:2013p26313,Schmidt:2014p33431}.

In contrast, spectroscopic confirmation has been much harder to
achieve. In spite of numerous attempts on modern and sensitive
ground-based spectrographs only very few \lya\ detections have
been reported at $z\sim7$ \citep[e.g.,][]{Pentericci:2011p27723,bradac2012,Ono:2012p27651,2012ApJ...744..179S,Capak:2013p31627,Caruana:2013p32713,Treu:2013p32132} and only one at
$z\gtrsim7.5$ \citep[][at $z=7.51$]{Finkelstein:2013p32467}. The
decrease in \lya\ flux coming from Lyman Break Galaxies (LBGs) at
$z\gtrsim6$ has been interpreted as evidence for an increased hydrogen
neutral fraction corresponding to a long ending tail of cosmic
reionization
\citep[e.g.,][]{Fontana:2010p29506,2013MNRAS.428.1366J,2013MNRAS.tmp.2740T}.
Alternative explanations for the decrease in \lya\ emission include
changes in the local circumgalactic medium \citep[e.g.,][]{2007MNRAS.377.1175D,2012ApJ...751...51J}
or perhaps a dramatic increase in the fraction of interlopers amongst
Lyman break galaxies.  Sensitive and complete spectroscopic surveys
sensitive to \lya\ at $z\gtrsim6$ are needed to make progress.

A powerful alternative to ground based spectroscopy is slitless
spectroscopy with \HST. Basic advantages over the ground are the
absence of night sky emission lines and of atmospheric
absorption. Furthermore, slitless spectroscopy does not require
preselection of targets for inclusion in masks and it is therefore
straightforward to obtain complete samples. These two advantages make
Hubble competitive for the spectroscopic study of galaxies at the
epoch of cosmic reionization provided that long enough exposures are
obtained, assisted by lensing magnification
\citep{Treu:2012p12658}.

Studying the very high-redshift universe with the combination of the
\HST\ grism spectroscopy and gravitational lensing magnification is one
of the key goals of the {\it Grism Lens-Amplified Survey from Space}
(GLASS), which we introduce in this letter.  We briefly present the
Survey, its goals and observational strategy and Wide Field Camera 3
(WFC3) infrared observations of the first targeted cluster \M. In
order to demonstrate the performance of the \HST\ grism in a crowded
field, we concentrate on spectroscopy of faint targets, including
multiply-imaged emission-line galaxies and photometrically-selected
galaxy candidates at $z\gtrsim6$. We also present additional ground
based spectroscopy obtained with the MOSFIRE
spectrograph on the W.M.~Keck-I 10m~Telescope.

We adopt a standard cosmology with $\Om=0.3$, $\OL=0.7$, $h=0.7$.

\section{The Grism Lens-Amplified Survey from Space}
\label{sec:glass}

GLASS\footnote{\url{http://glass.physics.ucsb.edu}} (GO-13459; PI: Treu) 
is a 140 orbit spectroscopic survey with
\HST.  Using WFC3's G102 and G141 infrared grisms GLASS will
obtain slitless spectroscopy of the cores of 10 galaxy clusters, with
uninterrupted wavelength coverage in the wavelength range 0.8--1.7
$\mu$m. Pre-imaging through filters F105W and F140W is obtained before
each spectral exposure to assist in the extraction of the spectra and
modeling the contamination from neighboring sources. The total
exposure time per cluster is 10 orbits in F105W+G102 and 4 in
F140W+G141. Each cluster is observed at two orientations, differing by
approximately 90 degrees, in order to facilitate deblending and
extraction of the spectra. Parallel observations with the Advanced
Camera for Surveys (ACS) through filter F814W and grism G800L are
carried out for each cluster in order to map the cluster infall
regions. The three key goals of GLASS are i) the study of galaxies at
the epoch of reionization through the detection of \lya\ at
$z\gtrsim6$; ii) the study of the cycling of gas and metals in and out
of galaxies at $2 < z < 4$; and iii) the study of star formation and
metallicity as a function of environment in the foreground clusters
and infall regions. Magnification by the foreground clusters greatly
assists the first two science goals as well as a number of other
ancillary science goals (e.g. lensed supernovae, red quiescent
galaxies, etc.).

\section{Observations and Data Reduction}
\label{sec:datar}

The GLASS data presented provide the first orientation (i.e., half of
the final data) of the cluster \M{} and were carried out on December
24 and 30 2013. The total exposure times are 10029 and
3812 seconds with the WFC3 G102 and G141 grisms, after removal of a
few reads affected by significantly elevated backgrounds (Brammer et
al. 2014, in prep.). In addition, we obtain 1979 and 712 seconds of
direct imaging in the WFC3 F105W and F140W filters, aiding the
alignment of the grism exposures.

\begin{figure*}
\begin{center}
\includegraphics[width=0.9\textwidth]{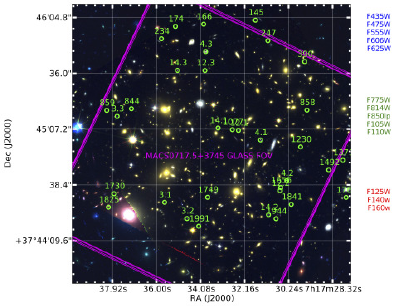}\\
\includegraphics[width=0.45\textwidth]{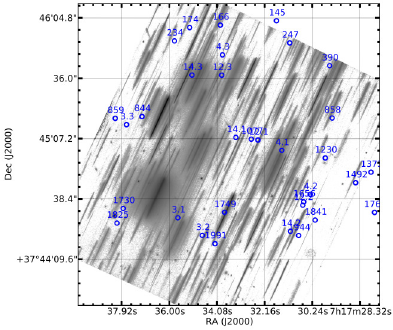}
\includegraphics[width=0.45\textwidth]{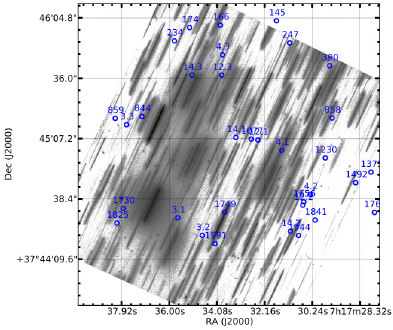}
\caption{The top panel shows a color composite image of \M\ based on the CLASH \citep{Postman:2012p27556} \HST\ data. The blue, green and red channels are composed by the filters on the right.
The magenta squares show the FOV of the GLASS pointings presented in this letter. 
The bottom panels show the GLASS G102 (left) and G141
(right) grism images of \M.  The location of the dropout candidates
and the multiple imaged sources from Table~\ref{tab:dropouts} are
indicated by the green (top panel) and blue (bottom panels) circles.
}
\label{fig:image}
\end{center}
\end{figure*}

\M{} is part of the Frontier Fields initiative\footnote{http://www.stsci.edu/hst/campaigns/frontier-fields/} and has been observed extensively by CLASH \citep{Postman:2012p27556}. We take advantage of the CLASH photometry when selecting our LBG samples as described in Section~\ref{sec:LBG}. In the top panel of Figure~\ref{fig:image} we show a false-color image using the CLASH imaging of \M{} with the position of the GLASS \M{} field-of-view (FOV) presented here marked by the magenta square. 

The GLASS observations were designed to follow the 3D-HST observing
strategy \citep{Brammer:2012p12977}, and were processed with
the 3D-HST reduction
pipeline\footnote{http://code.google.com/p/threedhst/}.  In the
following we summarize the main reduction steps but refer to
\cite{Brammer:2012p12977} for further details.

The data were taken in a 4-point dither pattern identical to the one
shown in Figure~3 of \cite{Brammer:2012p12977}.  At each individual
dither position a F105W-G102 or F141W-G141 pair of exposures was
taken, in order to optimize rejection of bad pixels and cosmic rays
and improve sampling of the WFC3 point spread function.

The individual exposures were turned into a mosaic using MultiDrizzle
\citep{Koekemoer:2003p31861} adjusting the alignments using the PyRAF
routine, \verb+tweakshifts+.  The background was subtracted from the
direct images fitting a 2nd order polynomial to each of the
source-subtracted exposures.  For the G102 background subtraction we
used the master background presented by \cite{Kummel:2011p33451},
whereas for the G141 grism we used the master backgrounds developed by
\cite{Brammer:2012p12977} for 3D-HST.

The individual background-subtracted exposures were then combined to
produce a mosaic with a final pixel scale of roughly $0\farcs06$ per
pixel as described by \citet{Brammer:2013p27911} ($\sim$half a native pixel). 
This corresponds to $\sim$12(22)\AA{} per pixel for the G102(G141) dispersion.
In the bottom panels of Figure~\ref{fig:image} we show the resulting interlaced full FOV
G102 (left) and G141 (right) grism images.

From these interlaced images the position of each individual spectrum
can be predicted by dispersing each pixel of each individual object in
the \verb+SExtractor+ \citep{Bertin:1996p12964} segmentation maps
created from the direct F105W and F140W images. In this way,
individual spectra can be extracted and spectral contamination, i.e.,
flux from overlapping spectra from neighboring objects, can be
accounted for.

\subsection{Additional ground based spectroscopy}

\M{} was observed at Keck on 2013 December 15(17) in Y(H) band
using the MOSFIRE near-infrared spectrograph \citep[][Keck Proposal U004M; PI Brada\v{c}]{McLean:2012p26812}. 
Conditions on December 15(17) were good(poor) with $0\farcs7$($1\farcs5$) seeing.
Exposure time was 3960s in Y and 3360s in H band. The data reduction
was performed using the publicly available MOSFIRE data reduction
pipeline (DRP\footnote{\url{http://code.google.com/p/mosfire/}}).

\section{Faint multiply-imaged galaxies}
\label{sec:mul}

\begin{figure*}
\begin{center}
\includegraphics[width=0.95\textwidth]{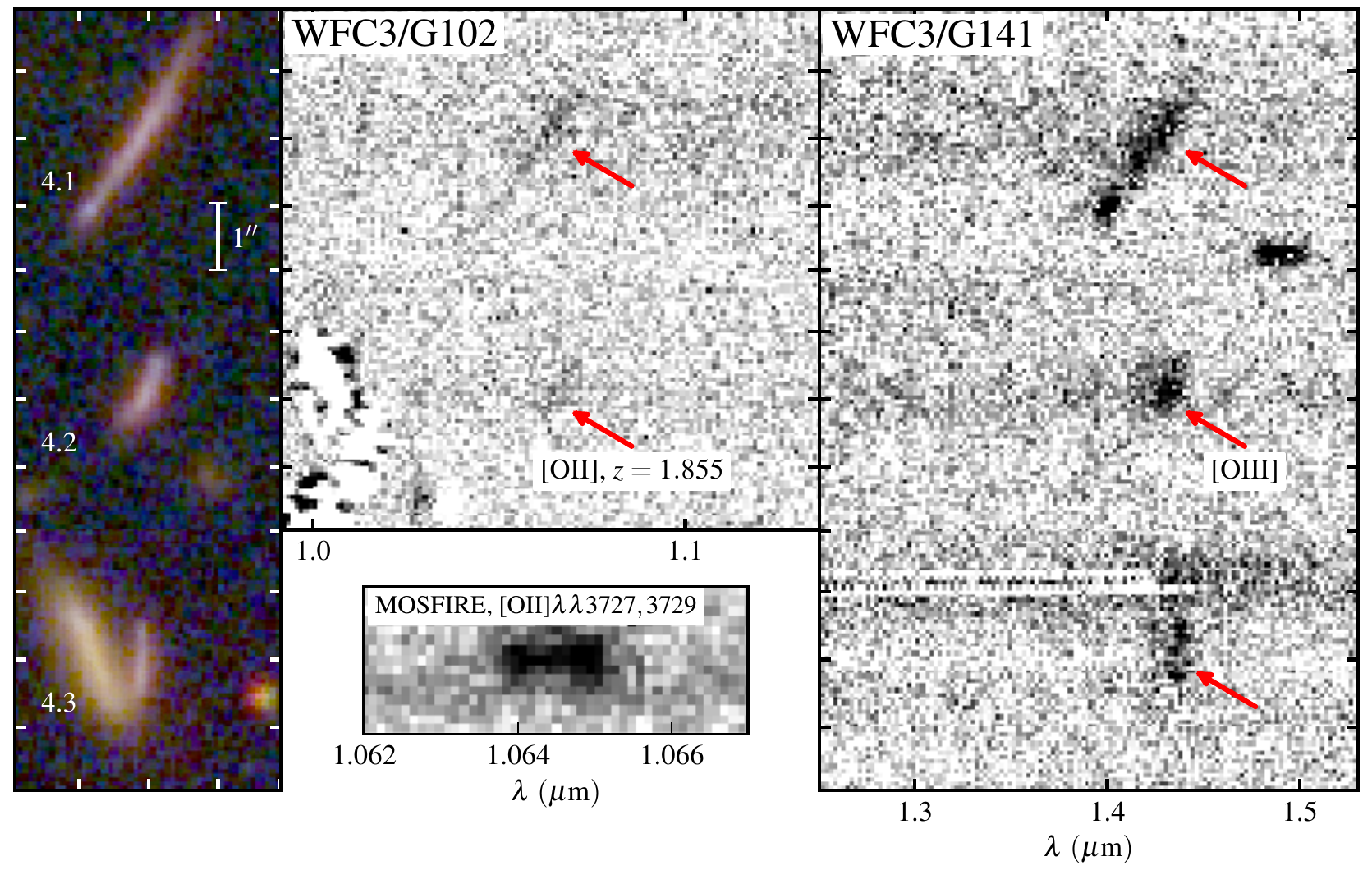}
\caption{Postage stamp images and contamination-subtracted spectra of the multiply imaged system 4 from Table~\ref{tab:dropouts} with detected emission lines (marked by red arrows). The G102 spectrum of 4.3 falls outside the GLASS FOV, but the line is clearly detected in ground based MOSFIRE observations as shown. Spatial ticks show 1$\arcsec$ intervals. For objects 3, 12, and 14 see Figure~\ref{fig:arcsG141}.}
\label{fig:arcsOBJ4}
\end{center}
\end{figure*}

\begin{figure}
\begin{center}
\includegraphics[width=0.40\textwidth]{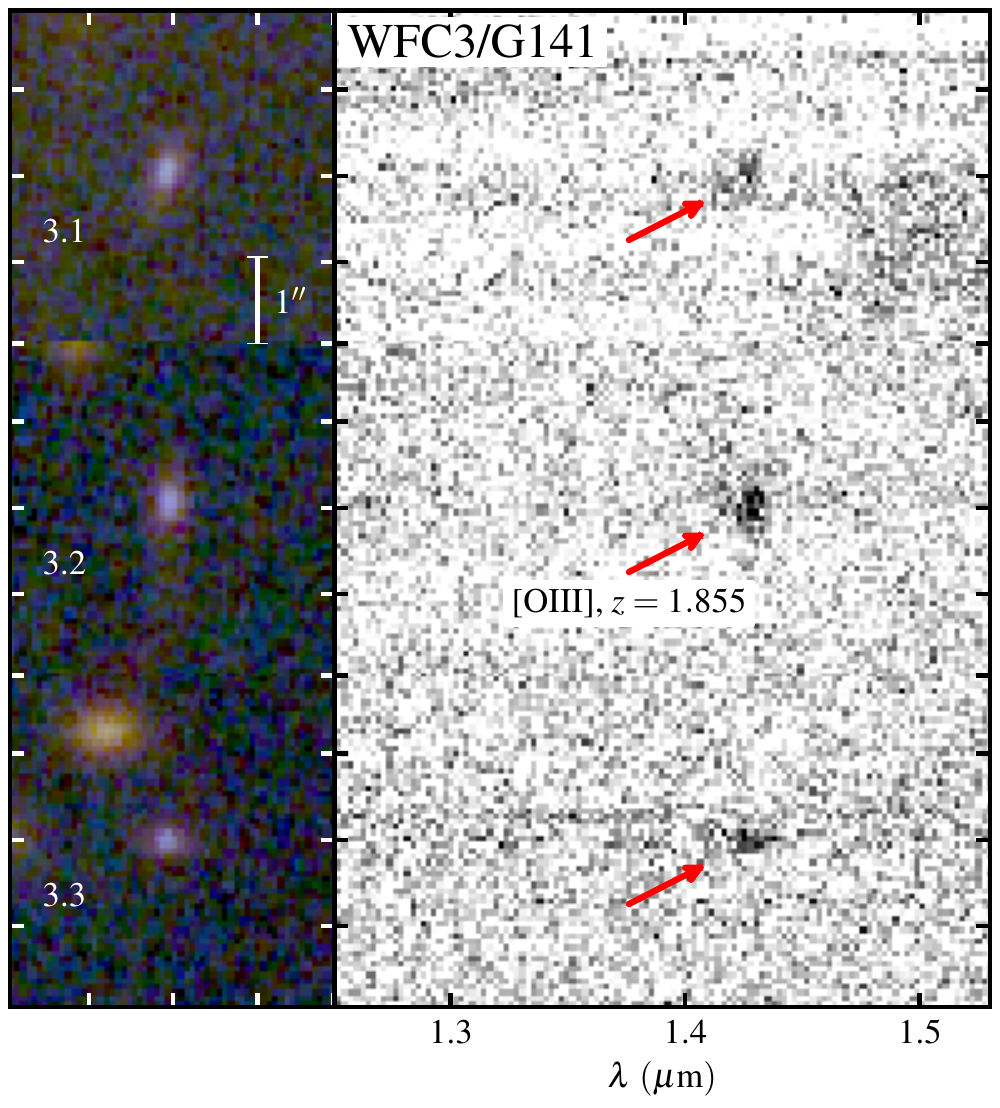}\\
\includegraphics[width=0.40\textwidth]{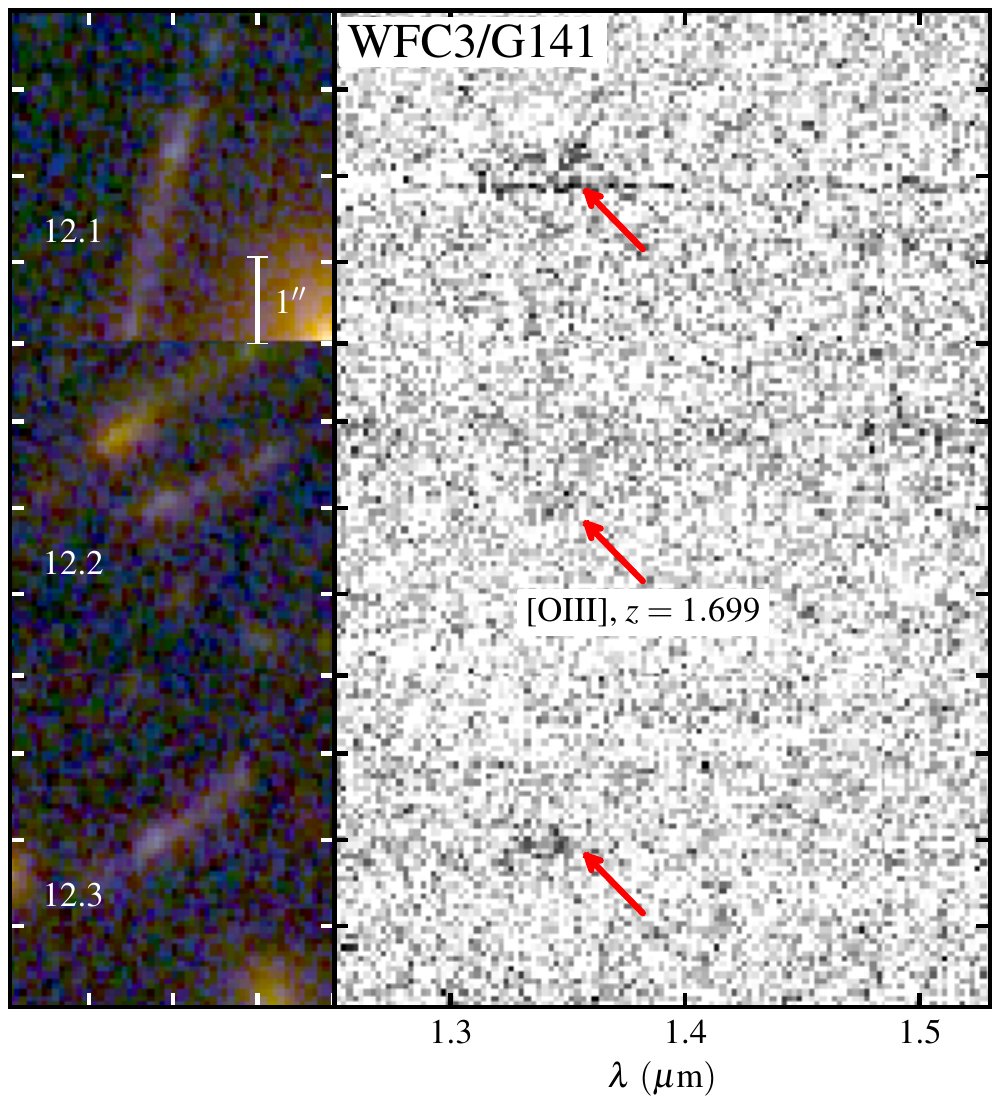}\\
\includegraphics[width=0.40\textwidth]{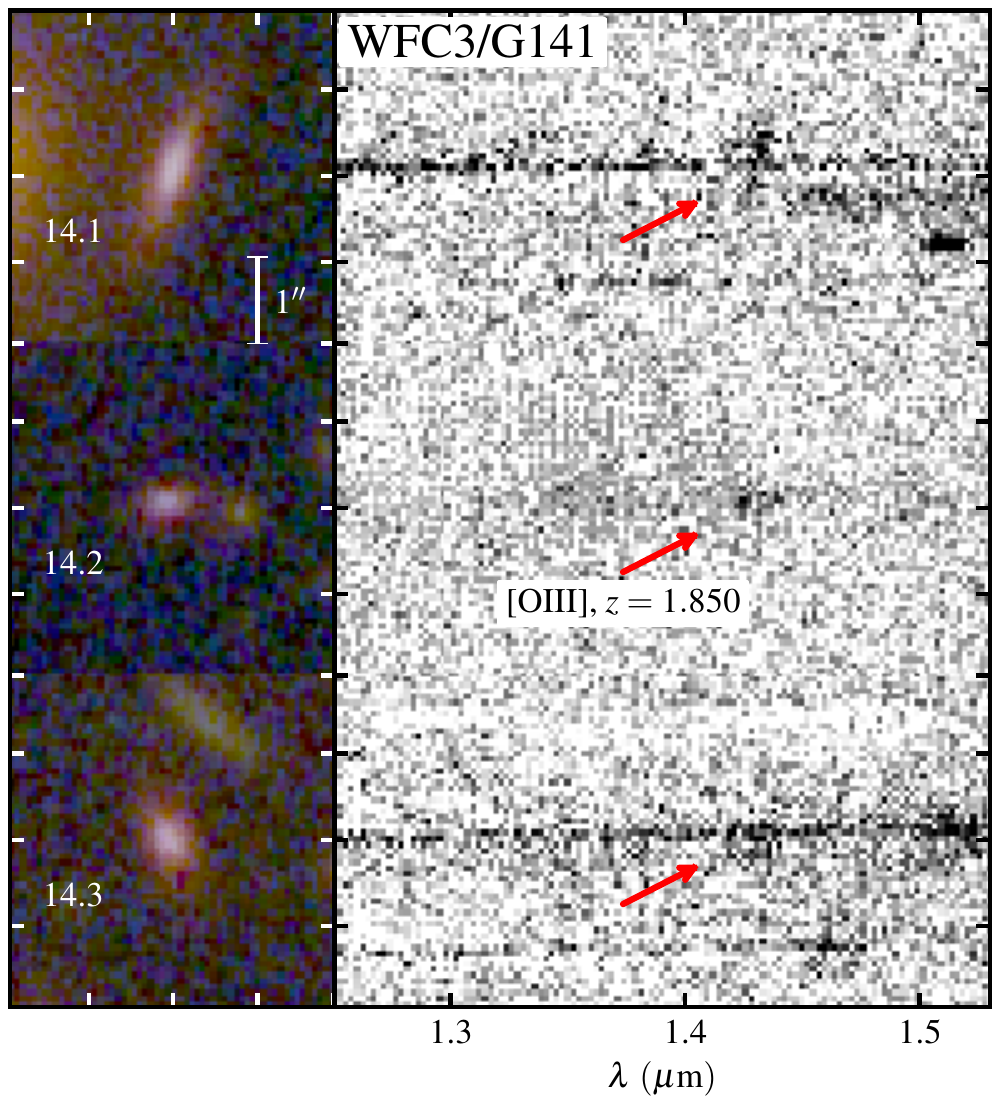}
\caption{Similar to Figure~\ref{fig:arcsOBJ4} for the multiply imaged systems 3, 12, and 14.}
\label{fig:arcsG141}
\end{center}
\end{figure}

We extracted spectra for all the sets of multiple images identified by
\citet{2012A&A...544A..71L} \citep[see also][for additional models of the
cluster]{2009ApJ...707L.102Z,2013ApJ...777...43M} that lie within
the GLASS FOV and do not have complete sets of
redshifts from the literature. For sets 3, 4, 12, 14 as defined by
\citet{2012A&A...544A..71L}, we detect the same emission
lines from each image (see Table~\ref{tab:dropouts}), thus confirming the lensing
hypothesis. The postage stamp images and two-dimensional spectra of
the four systems are shown in Figures~\ref{fig:arcsOBJ4} and \ref{fig:arcsG141}.  
Contaminating spectra from nearby objects have been modeled and subtracted from these spectra.
MOSFIRE spectra are shown together with GLASS \HST\ spectra when available.  
We also detect with MOSFIRE the [\ion{O}{3}]4959,5007 doublet in emission from image
6.1, which falls outside of the GLASS FOV, placing it at
$z=2.393$. The redshifts of image sets 3 and 14 agree with the value
for 3.2 and 14.1 ($z=1.85$) previously measured by
\citet{2012A&A...544A..71L}. 

Note the spatial extent of the emission lines, e.g. for 4.1. 
In such cases the combination of \HST's angular resolution and lensing 
magnification will allow us to measure spatially resolved metallicity and star 
formation gradients with sub-kpc resolution (Jones et al. 2014, in prep.).

The confirmation of the lensing hypothesis and the spectroscopic
redshifts being close to the photometric redshift estimates
($2.0\pm0.1$ $2.1\pm0.1$ $1.8\pm0.1$, respectively for sets 4,~6,~and
12) further validates the lens models published and available as part
of the Frontier Fields initiative. When the GLASS dataset of \M\ is
complete we will carry out a systematic search for new sets of
multiple image systems based on spectroscopic data. The use of
spectroscopic data greatly reduces the space of possible counterimages
and therefore should enable the unambiguous identification of faint
multiple images.  In combination with upcoming deep imaging from the
Frontier Fields initiative, the additional data will allow us to further refine the
lens models, the magnification maps, and study the distribution of
luminous and dark matter in the cluster itself.

    \tabletypesize{\tiny} \tabcolsep=0.11cm
    \begin{deluxetable*}{lccccccccccclr} \tablecolumns{15}
    \tablewidth{0pt} \tablecaption{Multiply-imaged and $z\gtrsim6$ galaxies in \M{}}
    \tablehead{\colhead{ID} & \colhead{$\alpha_\textrm{J2000}$} &
    \colhead{$\delta_\textrm{J2000}$} & \colhead{F814W mag} &
    \colhead{F850LP mag} & \colhead{F105W mag} & \colhead{F110W mag} &
    \colhead{F125W mag} & \colhead{F140W mag} & \colhead{F160W mag} & \colhead{$\mu$} &
    Line; C & $z$; Sel. & $f_\textrm{line}$; $\sigma_W$ }
\startdata
\hline
 3.1      & 0.398545 &  0.741498   &  24.99$\pm$0.04        &    24.88$\pm$0.09        &    25.05$\pm$0.07        &    25.02$\pm$0.05        &    24.93$\pm$0.06        &    24.61$\pm$0.04        &    24.50$\pm$0.04     &     4.6[5.1-20.0]             &  [OIII]           &  1.855       & 4.97$\pm$1.25  \\ 
 3.2      & 0.394459 &  0.739172   &  25.13$\pm$0.04        &    24.95$\pm$0.11        &    25.02$\pm$0.06        &    25.03$\pm$0.05        &    24.99$\pm$0.07        &    24.61$\pm$0.04        &    24.48$\pm$0.03     &     15.1[5.9-15.6]           &  [OIII]           &  1.855      & 11.89$\pm$1.19  \\ 
 3.3      & 0.407156 &  0.753831   &  25.77$\pm$0.07        &    25.99$\pm$0.23        &    25.69$\pm$0.10        &    25.51$\pm$0.07        &    25.34$\pm$0.08        &    25.23$\pm$0.06        &    25.05$\pm$0.05     &     3.1[3.1-6.2]               &  [OIII]           &  1.855       & 2.39$\pm$1.24  \\ 
 4.1      & 0.381093 &  0.750440   &  22.99$\pm$0.01        &    22.88$\pm$0.03        &    22.57$\pm$0.01        &    22.45$\pm$0.01        &    22.32$\pm$0.01        &    22.13$\pm$0.01        &    22.04$\pm$0.01     &     4.0[5.1-14.6]             &  [OIII]           &  1.855      & 46.12$\pm$2.42   \\ 
 4.2      & 0.376338 &  0.744602   &  23.84$\pm$0.02        &    23.68$\pm$0.05        &    23.45$\pm$0.02        &    23.30$\pm$0.01        &    23.20$\pm$0.02        &    23.02$\pm$0.01        &    22.94$\pm$0.01     &     9.5[3.1-9.5]               &  [OIII]           &  1.855      & 22.10$\pm$1.61     \\ 
 4.3      & 0.391097 &  0.763077   &  21.56$\pm$0.00        &    21.36$\pm$0.01        &    21.28$\pm$0.01        &    21.22$\pm$0.00        &    21.16$\pm$0.01        &    21.02$\pm$0.00        &    20.97$\pm$0.00     &     3.0[2.6-3.2]               &  [OIII]           &  1.855      & 21.61$\pm$1.38      \\ 
12.1     & 0.385131 &  0.751820   &  24.77$\pm$0.04        &    24.79$\pm$0.13        &    24.31$\pm$0.05        &    24.18$\pm$0.03        &    24.01$\pm$0.04        &    24.09$\pm$0.03        &    24.19$\pm$0.04     &      15.8[3.9-7.3]            &  [OIII]           &  1.699       & 4.15$\pm$1.04      \\ 
12.2     & 0.377607 &  0.742896   &  25.70$\pm$0.07        &    26.18$\pm$0.31        &    25.39$\pm$0.09        &    25.31$\pm$0.06        &    24.94$\pm$0.07        &    25.11$\pm$0.06        &    25.09$\pm$0.06     &      7.6[2.7-7.6]              &  [OIII]           &  1.699       & 4.40$\pm$1.13     \\ 
12.3     & 0.391226 &  0.760674   &  25.14$\pm$0.05        &    25.08$\pm$0.15        &    24.78$\pm$0.06        &    24.53$\pm$0.04        &    24.39$\pm$0.05        &    24.50$\pm$0.04        &    24.62$\pm$0.05     &      3.3[3.3-6.4]              &  [OIII]           &  1.699       & 8.98$\pm$1.35     \\ 
14.1     & 0.388806 &  0.752160   &  23.26$\pm$0.02        &    23.14$\pm$0.05        &    22.79$\pm$0.02        &    22.63$\pm$0.01        &    22.50$\pm$0.02        &    22.34$\pm$0.01        &    22.27$\pm$0.01     &      16.8[5.7-12.2]          &  [OIII]           &  1.855      &  5.32$\pm$1.76      \\ 
14.2     & 0.379659 &  0.739712   &  25.28$\pm$0.05        &    24.93$\pm$0.11        &    24.73$\pm$0.05        &    24.65$\pm$0.04        &    24.48$\pm$0.05        &    24.31$\pm$0.03        &    24.22$\pm$0.03     &      6.4[2.6-5.3]              &  [OIII]           &  1.855      &  4.61$\pm$1.24       \\ 
14.3     & 0.396192 &  0.760425   &  24.15$\pm$0.52        &    24.00$\pm$0.07        &    23.62$\pm$0.03        &    23.48$\pm$0.02        &    23.25$\pm$0.02        &    23.16$\pm$0.02        &    23.04$\pm$0.01     &      3.0[4.2-5.4]              &  [OIII]           &  1.855      &  2.18$\pm$1.67      \\ 
\hline
145               &   0.382022 &   0.767622 &     27.62$\pm$0.20    &   26.28$\pm$0.18       &   26.93$\pm$0.20      &   26.77$\pm$0.10     &    --                             &    26.91$\pm$0.20     &   27.09$\pm$0.22           &   13.8[2.8-13.8]           & o                  &  SB       &     --       \\
166               &   0.391443 &   0.767048 &     27.86$\pm$0.25    &   26.94$\pm$0.33       &   27.10$\pm$0.21      &   26.79$\pm$0.12     &    27.00$\pm$0.21      &   27.00$\pm$0.17      &   27.39$\pm$0.24           &   4.2[2.4-4.2]                & o                  & SB            &     --      \\
174               &   0.396578 &   0.766722 &     32.45$\pm$2.59    &   $>$28.63                  &   27.70$\pm$0.19      &  28.33$\pm$0.24     &    27.62$\pm$0.21      &   29.22$\pm$0.57      &   28.62$\pm$0.38            &   4.2[2.2-3.8]              & o                   &  zB              &     --       \\
234               &   0.399120 &   0.764958 &     28.06$\pm$0.39    &   26.33$\pm$0.25       &   26.45$\pm$0.15      &   26.11$\pm$0.08     &    26.49$\pm$0.18      &   26.11$\pm$0.10      &   26.34$\pm$0.13           &   4.2[2.8-4.2]                & g                  & iB,SB            &     43      \\
247\tnm{a}   &   0.379770 &   0.764690 &     26.70$\pm$0.13    &   25.87$\pm$0.16       &   25.58$\pm$0.06      &   25.58$\pm$0.05     &    25.46$\pm$0.07      &   25.42$\pm$0.05      &   25.43$\pm$0.05            &   9.5[4.0-22.8]             & o                  & SB              &    --       \\
390               &   0.373082 &   0.761676 &     28.43$\pm$0.29    &   27.88$\pm$0.51       &   27.26$\pm$0.17      &   27.42$\pm$0.14     &    27.47$\pm$0.22      &   27.54$\pm$0.19      &   27.39$\pm$0.17           &   3.9[3.1-8.9]                & o                  & SB            &     --      \\
844               &   0.404579 &   0.754928 &     27.20$\pm$0.15    &   26.63$\pm$0.28       &   26.47$\pm$0.14      &   26.53$\pm$0.11     &    26.63$\pm$0.17      &   26.69$\pm$0.14      &   26.78$\pm$0.16           &   7.4[7.4-28.1]              & c                  & SB            &     --      \\
858               &   0.372670 &   0.754728 &     27.85$\pm$0.21    &   27.11$\pm$0.32       &   27.16$\pm$0.18      &   27.48$\pm$0.17     &    27.25$\pm$0.21      &   27.20$\pm$0.16      &   27.61$\pm$0.23           &   28.5[3.9-28.5]            & g                  & iV,iB          &    117       \\
859               &   0.409065 &   0.754686 &     28.04$\pm$0.26    &   26.24$\pm$0.16       &   26.42$\pm$0.11      &   26.31$\pm$0.08     &    26.40$\pm$0.12      &   26.52$\pm$0.11      &   26.89$\pm$0.15           &   5.4[5.4-12.0]              & g                  & iV,SB        &     50      \\
1077             &   0.386225 &   0.751928 &     28.36$\pm$0.36    &   27.01$\pm$0.31       &   26.92$\pm$0.17      &   27.00$\pm$0.14     &    27.13$\pm$0.22      &   26.98$\pm$0.16      &   27.02$\pm$0.17           &   2.5[2.4-3.1]                & g                  & SB            &     84      \\
1230\tnm{a} &   0.373824 &   0.749445 &     26.69$\pm$0.12    &   25.87$\pm$0.16       &   25.63$\pm$0.07      &   25.54$\pm$0.05     &    25.68$\pm$0.08      &   25.48$\pm$0.05      &   25.56$\pm$0.06           &   34.4[9.7-34.4]            & g                  & SB             &     21      \\
1373             &   0.366140 &   0.747541 &     28.37$\pm$0.28    &   29.18$\pm$1.16       &   27.94$\pm$0.30      &   27.33$\pm$0.14     &    27.54$\pm$0.24      &   27.75$\pm$0.23      &   27.43$\pm$0.18           &   1.8[1.4-1.8]                & o                  & SB            &     --      \\
1492             &   0.368709 &   0.746143 &     27.59$\pm$0.18    &   27.00$\pm$0.32       &   27.66$\pm$0.31      &   27.08$\pm$0.14     &    27.07$\pm$0.21      &   27.16$\pm$0.18      &   27.15$\pm$0.18           &   13.1[3.1-13.1]            & c                  & iB              &      --     \\
1656             &   0.377434 &   0.743617 &     28.97$\pm$0.53    &   $>$28.23                  &   27.38$\pm$0.24      &  27.78$\pm$0.25     &    27.25$\pm$0.24      &   27.25$\pm$0.19      &   27.15$\pm$0.18            &   36.5[5.2-36.5]          & g                  & zB               &     74       \\
1730             &   0.407728 &   0.742741 &     28.43$\pm$0.32    &   25.93$\pm$0.11       &   26.35$\pm$0.16      &   --                            &    26.79$\pm$0.16       &   26.71$\pm$0.16     &   26.78$\pm$0.18           &   2.8[4.7-18.3]             & c                  & iV,SB          &     --       \\
1749             &   0.390730 &   0.742226 &     26.28$\pm$0.11    &   25.71$\pm$0.21       &   25.30$\pm$0.08      &   25.29$\pm$0.06     &    25.44$\pm$0.10      &   25.46$\pm$0.09      &   25.41$\pm$0.08           &   4.6[12.9-39.0]            & c                  & SB            &     --      \\
1764             &   0.365558 &   0.742210 &     28.53$\pm$0.34    &   27.24$\pm$0.31       &   27.37$\pm$0.21      &   27.64$\pm$0.19     &    27.04$\pm$0.19      &   27.61$\pm$0.23      &   27.63$\pm$0.23           &   5.6[1.9-5.6]                & o                  & iB              &     --      \\
1825             &   0.408768 &   0.740805 &     25.49$\pm$0.03    &   24.69$\pm$0.04       &   24.82$\pm$0.04      &   --                            &    24.86$\pm$0.03       &   24.87$\pm$0.03     &   24.82$\pm$0.03           &   5.0[5.0-32.7]             & c                  & SB             &      --      \\
1841             &   0.375517 &   0.741206 &     28.06$\pm$0.19    &   27.21$\pm$0.26       &   28.17$\pm$0.33      &   28.01$\pm$0.22     &    27.67$\pm$0.25      &   28.02$\pm$0.26      &   28.22$\pm$0.31           &   6.3[2.7-6.3]                & g                  & iB              &     197      \\
1944             &   0.378288 &   0.739163 &     27.72$\pm$0.21    &   26.41$\pm$0.20       &   27.50$\pm$0.28      &   27.31$\pm$0.18     &    27.01$\pm$0.21      &   26.99$\pm$0.16      &   27.21$\pm$0.19           &   6.4[2.9-6.4]                & g                  & iB,SB        &    92       \\
1991             &   0.392335 &   0.738083 &     26.05$\pm$0.09    &   25.55$\pm$0.24       &   25.47$\pm$0.09      &   25.66$\pm$0.08     &    25.54$\pm$0.11      &   25.50$\pm$0.08      &   25.59$\pm$0.09           &   36.5[8.7-36.5]            & g                  & iV,iB,SB    &     23      \\
\enddata   
\tablecomments{Multiply-imaged (Section~\ref{sec:mul}) and $z\gtrsim6$ 
(Section~\ref{sec:LBG}) galaxy samples.  Right ascension
$\alpha_\textrm{J2000}$ is relative to 109.000000 and declination
$\delta_\textrm{J2000}$ is relative to 37.000000.  Magnitudes are
observed CLASH (lensed) isophotal magnitudes (ISOMAG).  The column
$\mu$ gives the best fit magnification estimate at the object position
from the Frontier Fields magnification map of Bradac et al. (see
http://archive.stsci.edu/prepds/frontier/lensmodels/). The range after
each value shows the minimum an maximum estimate of seven different
Frontier Fields magnification estimates after removing the largest and
smallest values.  For seven models this approximates the 16th and 84th
percentile of the distribution, and thus provides an estimate of the
modeling error.  The remaining three columns have different meaning
for the two samples.  For the multiple imaged sources they indicate
the main emission line found in the grism spectra (line), the corresponding
redshift ($z$), and the estimated line flux ($f_{\rm line}/[\textrm{1e-17 erg/s/cm}^2]$). 
For the dropouts C indicates whether an object fell outside the GLASS FOV
shown in Figure~\ref{fig:image} (o), was heavily contaminated (c), or
is considered good enough for EW limit estimates (g). The Sel. column
indicates what selection criteria each object satisfies as described
in Section~\ref{sec:sel}, and $\sigma_W$ lists the 1$\sigma$ noise
level for each target expressed in rest frame equivalent width.  In
order to compute $\mu$ and $\sigma_W$ we adopt fiducial redshifts
$z=6$ for SB,~iV,~iB, $z=7$ for zB, and $z=6.4$ for the two
spectroscopically confirmed sources (859, 1730).
\tnm{a}{Unresolved object with FWHM $<0.22\arcsec$ in the image plane \citep[see][]{Bradley:2013p32053}}
}
\label{tab:dropouts}
\end{deluxetable*}

\section{Galaxy candidates at $z>6$}
\label{sec:LBG}

\subsection{Photometric Selection}\label{sec:sel}

We have assembled an extensive list of candidate galaxies at
$z\sim6-8$ using multiple selection criteria applied to the CLASH
catalog as summarized below.  After the initial color/SED criteria,
all candidates were visually inspected to remove hot pixels,
diffraction spikes, and edge defects. The candidates passing the color
selection and visual inspection are listed in
Table~\ref{tab:dropouts}.

\begin{itemize}
\item[iV] i$_{775}$-band dropout selection from \cite{Vanzella:2009p29479}, 
requiring that all bands bluewards of the F775W band have
S/N$<$2, i$-$z$>$1.3 and S/N$_\textrm{z}>5$. 
This selection yields 4 candidates with $5.5\lesssim z\lesssim6.5$.

\item[iB] i$_{775}$-band dropout selection from \cite{Bouwens:2012p10416}. 
Similar to iV except that z$-$J$<$0.9 instead of S/N$_\textrm{z}>5$.  This
selection yields 7 candidates with $5.5\lesssim z\lesssim6.5$.

\item[zB] z$_{850}$-band dropout selection from \cite{Bouwens:2012p10416}. 
Bands bluewards of F850LP were required to have S/N$<2$ and
\begin{eqnarray}
\textrm{z} -\textrm{Y} &>& 0.7 \nonumber \\
\textrm{Y} -\textrm{J}  &<& 0.8 \nonumber \\
\textrm{z} - \textrm{Y} &>& 1.4\times (\textrm{Y} - \textrm{J}) + 0.42
\end{eqnarray}
This returns 2 candidates with $6.0\lesssim z\lesssim8.0$.

\item[SB] The 15 CLASH SED-selected $z\sim6$ LBGs behind \M{} from \cite{Bradley:2013p32053}. 
\end{itemize}

Column `Sel.' of Table~\ref{tab:dropouts} indicates the selection criteria
that yielded each candidate. Note that some candidates are identified
by multiple criteria. We also searched for YJ, J$_{125}$, and
JH$_{140}$ dropouts as described by \cite{Oesch:2013p27877}, Y
dropouts following \cite{Bouwens:2011p8082}, and a slightly modified
(using F105W instead of F098M)
version of the BoRG $z\sim8$ Y-band dropouts selection
\citep{Trenti:2011p12656,Bradley:2012p23263,Schmidt:2014p33431}.  
None of these methods returned any candidates.  
The \cite{Bradley:2013p32053} photometric-redshift selection agrees with this
as they did not find any $z\sim7-8$ candidates either.
In summary, we have assembled a list of \Nhigh\
unique candidate high-$z$ galaxies. Two of the targets (859 and 1730)
have recently been spectroscopically confirmed to be at $z=6.4$ by
\citet{Vanzella:2013p33637}.

\subsection{Flux, luminosity and equivalent width limits.}

\begin{figure}
\includegraphics[width=0.49\textwidth]{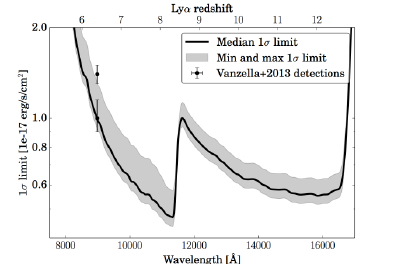}
\includegraphics[width=0.49\textwidth]{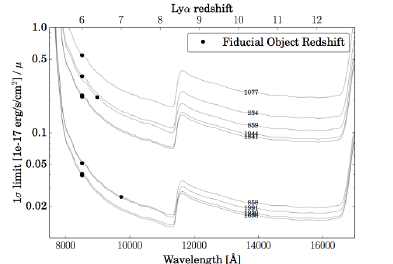}
\caption{The top panel shows 1$\sigma$ flux limits estimated from the \Nok\ 2D GLASS spectra (out of \Nspec\ spectra in total) found suitable for flux limit estimates, i.e., no contamination. The shaded region shows the spread in values. The two 15$\sigma$ and 9$\sigma$ \lya\ detections for object 859 and 1730 presented by \cite{Vanzella:2013p33637} are shown for reference (the uncertainty on the Ly$\alpha$ redshift is smaller than the extend of the dot). The bottom panel shows the \Nok\ individual estimates divided by the estimated magnification at each object position from the Bradac et al. Frontier Fields lensing model (see Table~\ref{tab:dropouts}). Each curve is marked by the ID from Table~\ref{tab:dropouts} and the objects fiducial redshift (Ly$\alpha$ wavelength). For the objects that will be observed with the second orientation, the sensitivity is going to improve by a factor of $\sim\sqrt{2}$ at the end of the GLASS survey.}
\label{fig:Flimit}
\end{figure}

Out of the \Nhigh\ LBG candidates \Nspec\ fell within the FOV of the
G102 and G141 GLASS pointings analyzed here.  We do not detect any
line emission. \Nok\ of the GLASS spectra were free of contamination
by spectra from other objects and were used to estimate the flux
sensitivity.\footnote{For the remaining objects the flux sensitivity
depends very strongly on the wavelength range and the level
of contamination.}  For an aperture of 5~(spatial) by 3~(spectral)
native pixels ($\sim$0.6\arcsec$\times$72(132)\AA{} for G102(G141)), 
the 1$\sigma$ limiting flux is of the order
10$^{-17}$\cgs\ as shown in the top panel of
Figure~\ref{fig:Flimit}. The sensitivity is comparable with what is
reached in blank fields \citep[e.g.,][]{Atek:2010p33653} and agrees
well with the exposure time calculator.

One of the goals of GLASS is to take advantage of the lensing
magnification of high-redshift sources to enhance the probability of
detecting otherwise unreachable flux limits. To illustrate the
benefits of magnification, the bottom panel of Figure~\ref{fig:Flimit}
shows the limiting flux for the \Nok\ spectra corrected for the
estimated magnification, assuming that they are at a redshift of 6,
6.4, 7 depending on their selection method. Note that magnification
changes very little at these redshifts, owing to the flatness of the
angular diameter distance-redshift relation. It is clear that a
significant improvement in the effective limiting fluxes is achieved
from the cluster lensing magnification. However, it is also clear that
a relatively high magnification is needed to obtain several-$\sigma$
detections of line fluxes in the $\sim$10$^{-18}$\cgs\ regime, which
is the deepest practical limit for ground based optical and near-infrared studies of field galaxies
\citep[e.g.,][]{Stark:2011p27664,Treu:2013p32132}. 
Naturally, owing to the low and smooth background from space, the
sensitivity of GLASS is more or less independent of wavelength except
at the edges of the spectral coverage where the grisms throughput
drop.

Assuming the spectral energy distribution of the sources to be
approximately flat in $f_\nu$ over the G102 and G141 wavelength range
we can estimate the continuum flux for each source as a function of
wavelength from a weighted average of the CLASH F125W, F140W and F160W magnitudes.
In combination with the limiting flux the continuum flux gives the limiting equivalent width
for each of the \Nok\ objects with suitable spectra. The rest-frame
equivalent width limits at the fiducial redshift for each candidate
are given in Table~\ref{tab:dropouts}.

Given the intrinsic faintness of the candidates, the equivalent width
limits are rather high and it is thus not surprising that we do not
detect any \lya\ emission for this sample. Even at $z\sim6$ only
$\lesssim 4\%$ of Lyman Break Galaxies would be expected to have
emission larger than 100\AA\ and thus be significantly detected in our
exposures. However, the equivalent width limits are sufficient to rule
out pure line emitters at lower redshift as contaminants
\citep{Atek:2011p10369}.


\section{Conclusions}
\label{sec:conc}

We have introduced GLASS, an \HST\ Large Program devoted to the study
of the high-redshift universe by means of deep grism
spectroscopy. Using the first WFC3 pointing in the center of the rich
cluster \M, we have demonstrated the power of grism spectroscopy even
in very crowded fields. We have shown that by carefully modeling the
background contamination one can obtain deep spectra of large numbers
of sources at once, without the limitations due to preselection for
inclusion in masks or wavelength gaps typical of ground based
studies. In practice we have considerered two sets of faint targets
and obtained the following results:

\begin{enumerate}

\item We have confirmed \Nmul\ sets (3,4,12,14) of candidate strong gravitational 
lens systems by identifying the same emission lines in each of the
multiple images. Redshift for one additional system (6.1) was obtained from
ground based MOSFIRE spectroscopy. These 5 redshifts are close to the
photometric redshift estimates used to construct the lens models
provided by the Frontier Fields initiative. The confirmation of the
strong lensing hypothesis and redshift thus provides additional
validation of the models.

\item We have compiled a list of \Nspec\ photometrically selected candidate 
galaxies at $z\sim6-8$ that fall in the GLASS field of view. No \lya\
emission is detected.

\item We have used the spectra of the \Nok\ candidates with no significant
contamination from neighbouring objects to measure the line flux
sensitivity of GLASS and found it to be
0.5-1$\times$10$^{-17}$\cgs. This is consistent with the limits
obtained in blank fields. The completeness and sensitivity are going
to improve with the second set of visits, which will deliver the same
exposure time with a dispersion direction that is approximately
orthogonal to that of the data presented here.
\end{enumerate}

Even with just a fraction of the total observing time allocated for
GLASS, the observations illustrate the power of \HST\ grism spectroscopy
combined with strong gravitational lensing.  The complete dataset will
provide a treasure trove of spectroscopic information, which will be
useful to many in addressing several outstanding scientific questions.

\acknowledgments

This paper is based on observations made with the NASA/ESA Hubble
Space Telescope, obtained at STScI. We acknowledge support through
grants HST-13459, HST-GO13177, HST-AR13235. This work utilizes
gravitational lensing models produced by PIs Brada\v{c}, Ebeling,
Merten, Zitrin, Sharon, and Williams funded as part of the \HST\
Frontier Fields program conducted by STScI. STScI is operated by AURA,
Inc. under NASA contract NAS 5-26555. The lens models were obtained
from the Mikulski Archive for Space Telescopes (MAST).  The Dark
Cosmology Centre (DARK) is funded by the Danish National Research
Foundation. Some of the data presented herein were obtained at the
W.M. Keck Observatory. The authors wish to recognize and acknowledge
the very significant cultural role and reverence that the summit of
Mauna Kea has always had within the indigenous Hawaiian community.



\begin{thebibliography}{}
\expandafter\ifx\csname natexlab\endcsname\relax\def\natexlab#1{#1}\fi

\bibitem[{Atek {et~al.}(2010)Atek, Malkan, McCarthy, Teplitz, Scarlata, Siana,
  Henry, Colbert, Ross, Bridge, Bunker, Dressler, Fosbury, Martin, \&
  Shim}]{Atek:2010p33653}
Atek, H., Malkan, M., McCarthy, P., {et~al.} 2010, The Astrophysical Journal,
  723, 104

\bibitem[{Atek {et~al.}(2011)Atek, Siana, Scarlata, Malkan, McCarthy, Teplitz,
  Henry, Colbert, Bridge, Bunker, Dressler, Fosbury, Hathi, Martin, Ross, \&
  Shim}]{Atek:2011p10369}
Atek, H., Siana, B., Scarlata, C., {et~al.} 2011, eprint arXiv, 1109, 639

\bibitem[{Bennett {et~al.}(2013)Bennett, Larson, Weiland, Jarosik, Hinshaw,
  Odegard, Smith, Hill, Gold, Halpern, Komatsu, Nolta, Page, Spergel, Wollack,
  Dunkley, Kogut, Limon, Meyer, Tucker, \& Wright}]{Bennett:2013p27238}
Bennett, C.~L., Larson, D., Weiland, J.~L., {et~al.} 2013, The Astrophysical
  Journal Supplement, 208, 20

\bibitem[{Bertin \& Arnouts(1996)}]{Bertin:1996p12964}
Bertin, E., \& Arnouts, S. 1996, Astronomy and Astrophysics Supplement, 117,
  393

\bibitem[{Bouwens {et~al.}(2010)Bouwens, Illingworth, Oesch, Stiavelli, van
  Dokkum, Trenti, Magee, Labb{\'e}, Franx, Carollo, \&
  Gonzalez}]{Bouwens:2010p30142}
Bouwens, R.~J., Illingworth, G.~D., Oesch, P.~A., {et~al.} 2010, The
  Astrophysical Journal Letters, 709, L133

\bibitem[{Bouwens {et~al.}(2011)Bouwens, Illingworth, Oesch, Labb{\'e}, Trenti,
  van Dokkum, Franx, Stiavelli, Carollo, Magee, \&
  Gonzalez}]{Bouwens:2011p8082}
---. 2011, The Astrophysical Journal, 737, 90

\bibitem[{Bouwens {et~al.}(2012)Bouwens, Illingworth, Oesch, Franx, Labb{\'e},
  Trenti, van Dokkum, Carollo, Gonz{\'a}lez, Smit, \&
  Magee}]{Bouwens:2012p10416}
---. 2012, The Astrophysical Journal, 754, 83, 33 pages, 24 figures, 7 tables,
  submitted to ApJ

\bibitem[Brada{\v c} {et~al.}(2012)]{bradac2012} Brada{\v c}, M., 
Vanzella, E., Hall, N., et al.\ 2012, \apjl, 755, L7 

\bibitem[{Bradley {et~al.}(2012)Bradley, Trenti, Oesch, Stiavelli, Treu,
  Bouwens, Shull, Holwerda, \& Pirzkal}]{Bradley:2012p23263}
Bradley, L.~D., Trenti, M., Oesch, P.~A., {et~al.} 2012, The Astrophysical
  Journal, 760, 108

\bibitem[{Bradley {et~al.}(2013)Bradley, Zitrin, Coe, Bouwens, Postman,
  Balestra, Grillo, Monna, Rosati, Seitz, Host, Lemze, Moustakas, Moustakas,
  Shu, Zheng, Broadhurst, Carrasco, Jouvel, Koekemoer, Medezinski, Meneghetti,
  Nonino, Smit, Umetsu, Bartelmann, Benitez, Donahue, Ford, Infante,
  Jimenez-Teja, Kelson, Lahav, Maoz, Melchior, Merten, \&
  Molino}]{Bradley:2013p32053}
Bradley, L.~D., Zitrin, A., Coe, D., {et~al.} 2013, eprint arXiv, 1308, 1692

\bibitem[{Brammer {et~al.}(2013)Brammer, van Dokkum, Illingworth, Bouwens,
  Labb{\'e}, Franx, Momcheva, \& Oesch}]{Brammer:2013p27911}
Brammer, G.~B., van Dokkum, P.~G., Illingworth, G.~D., {et~al.} 2013, The
  Astrophysical Journal Letters, 765, L2

\bibitem[{Brammer {et~al.}(2012)Brammer, van Dokkum, Franx, Fumagalli, Patel,
  Rix, Skelton, Kriek, Nelson, Schmidt, Bezanson, Cunha, Erb, Fan, Schreiber,
  Illingworth, Labb{\'e}, Leja, Lundgren, Magee, Marchesini, McCarthy,
  Momcheva, Muzzin, Quadri, Steidel, Tal, Wake, Whitaker, \&
  Williams}]{Brammer:2012p12977}
Brammer, G.~B., van Dokkum, P.~G., Franx, M., {et~al.} 2012, The Astrophysical
  Journal Supplement, 200, 13

\bibitem[{Capak {et~al.}(2013)Capak, Faisst, Vieira, Tacchella, Carollo, \&
  Scoville}]{Capak:2013p31627}
Capak, P.~L., Faisst, A., Vieira, J.~D., {et~al.} 2013, eprint arXiv, 1307,
  4089, accepted to ApJL, 5 Pages, 4 Figures, 1 Table

\bibitem[{Caruana {et~al.}(2013)Caruana, Bunker, Wilkins, Stanway, Lorenzoni,
  Jarvis, \& Elbert}]{Caruana:2013p32713}
Caruana, J., Bunker, A.~J., Wilkins, S.~M., {et~al.} 2013, eprint arXiv, 1311,
  57, submitted to MNRAS

\bibitem[{Coe {et~al.}(2013)Coe, Zitrin, Carrasco, Shu, Zheng, Postman,
  Bradley, Koekemoer, Bouwens, Broadhurst, Monna, Host, Moustakas, Ford,
  Moustakas, Wel, Donahue, Rodney, Ben{\'\i}tez, Jouvel, Seitz, Kelson, \&
  Rosati}]{Coe:2013p26313}
Coe, D., Zitrin, A., Carrasco, M., {et~al.} 2013, The Astrophysical Journal,
  762, 32

\bibitem[{{Dijkstra} {et~al.}(2007){Dijkstra}, {Lidz}, \&
  {Wyithe}}]{2007MNRAS.377.1175D}
{Dijkstra}, M., {Lidz}, A., \& {Wyithe}, J.~S.~B. 2007, \mnras, 377, 1175

\bibitem[{Ellis {et~al.}(2012)Ellis, McLure, Dunlop, Robertson, Ono, Schenker,
  Koekemoer, Bowler, Ouchi, Rogers, Curtis-Lake, Schneider, Charlot, Stark,
  Furlanetto, \& Cirasuolo}]{Ellis:2012p26700}
Ellis, R.~S., McLure, R.~J., Dunlop, J.~S., {et~al.} 2012, eprint arXiv, 1211,
  6804

\bibitem[{Finkelstein {et~al.}(2013)Finkelstein, Papovich, Dickinson, Song,
  Tilvi, Koekemoer, Finkelstein, Mobasher, Ferguson, Giavalisco, Reddy, Ashby,
  Dekel, Fazio, Fontana, Grogin, Huang, Kocevski, Rafelski, Weiner, \&
  Willner}]{Finkelstein:2013p32467}
Finkelstein, S.~L., Papovich, C., Dickinson, M., {et~al.} 2013, eprint arXiv,
  1310, 6031

\bibitem[{Fontana {et~al.}(2010)Fontana, Vanzella, Pentericci, Castellano,
  Giavalisco, Grazian, Boutsia, Cristiani, Dickinson, Giallongo, Maiolino,
  Moorwood, \& Santini}]{Fontana:2010p29506}
Fontana, A., Vanzella, E., Pentericci, L., {et~al.} 2010, The Astrophysical
  Journal Letters, 725, L205

\bibitem[{Hinshaw {et~al.}(2013)Hinshaw, Larson, Komatsu, Spergel, Bennett,
  Dunkley, Nolta, Halpern, Hill, Odegard, Page, Smith, Weiland, Gold, Jarosik,
  Kogut, Limon, Meyer, Tucker, Wollack, \& Wright}]{Hinshaw:2013p27234}
Hinshaw, G., Larson, D., Komatsu, E., {et~al.} 2013, The Astrophysical Journal
  Supplement, 208, 19

\bibitem[{{Jensen} {et~al.}(2013){Jensen}, {Laursen}, {Mellema}, {Iliev},
  {Sommer-Larsen}, \& {Shapiro}}]{2013MNRAS.428.1366J}
{Jensen}, H., {Laursen}, P., {Mellema}, G., {et~al.} 2013, \mnras, 428, 1366

\bibitem[{{Jones} {et~al.}(2012){Jones}, {Stark}, \&
  {Ellis}}]{2012ApJ...751...51J}
{Jones}, T., {Stark}, D.~P., \& {Ellis}, R.~S. 2012, \apj, 751, 51

\bibitem[{Koekemoer {et~al.}(2003)Koekemoer, Fruchter, Hook, \&
  Hack}]{Koekemoer:2003p31861}
Koekemoer, A.~M., Fruchter, A.~S., Hook, R.~N., \& Hack, W. 2003, The 2002 HST
  Calibration Workshop : Hubble after the Installation of the ACS and the
  NICMOS Cooling System, 337

\bibitem[{K{\"u}mmel {et~al.}(2011)K{\"u}mmel, Kuntschner, Walsh, \&
  Bushouse}]{Kummel:2011p33451}
K{\"u}mmel, M., Kuntschner, H., Walsh, J.~R., \& Bushouse, H. 2011, ST-ECF
  Instrument Science Report WFC3-2011-01, 1

\bibitem[{{Limousin} {et~al.}(2012){Limousin}, {Ebeling}, {Richard},
  {Swinbank}, {Smith}, {Jauzac}, {Rodionov}, {Ma}, {Smail}, {Edge}, {Jullo}, \&
  {Kneib}}]{2012A&A...544A..71L}
{Limousin}, M., {Ebeling}, H., {Richard}, J., {et~al.} 2012, \aap, 544, A71

\bibitem[{McLean {et~al.}(2012)McLean, Steidel, Epps, Konidaris, Matthews,
  Adkins, Aliado, Brims, Canfield, Cromer, Fucik, Kulas, Mace, Magnone,
  Rodriguez, Rudie, Trainor, Wang, Weber, \& Weiss}]{McLean:2012p26812}
McLean, I.~S., Steidel, C.~C., Epps, H.~W., {et~al.} 2012, Ground-based and
  Airborne Instrumentation for Astronomy IV. Proceedings of the SPIE, 8446,
  doi:10.1117/12.924794

\bibitem[{{Medezinski} {et~al.}(2013){Medezinski}, {Umetsu}, {Nonino},
  {Merten}, {Zitrin}, {Broadhurst}, {Donahue}, {Sayers}, {Waizmann},
  {Koekemoer}, {Coe}, {Molino}, {Melchior}, {Mroczkowski}, {Czakon}, {Postman},
  {Meneghetti}, {Lemze}, {Ford}, {Grillo}, {Kelson}, {Bradley}, {Moustakas},
  {Bartelmann}, {Ben{\'{\i}}tez}, {Biviano}, {Bouwens}, {Golwala}, {Graves},
  {Infante}, {Jim{\'e}nez-Teja}, {Jouvel}, {Lahav}, {Moustakas}, {Ogaz},
  {Rosati}, {Seitz}, \& {Zheng}}]{2013ApJ...777...43M}
{Medezinski}, E., {Umetsu}, K., {Nonino}, M., {et~al.} 2013, \apj, 777, 43

\bibitem[{Oesch {et~al.}(2013)Oesch, Bouwens, Illingworth, Labb{\'e}, Franx,
  van Dokkum, Trenti, Stiavelli, Gonzalez, \& Magee}]{Oesch:2013p27877}
Oesch, P.~A., Bouwens, R.~J., Illingworth, G.~D., {et~al.} 2013, The
  Astrophysical Journal, 773, 75, 21 pages, 13 figures, 6 tables; submitted to
  ApJ

\bibitem[{Ono {et~al.}(2012)Ono, Ouchi, Mobasher, Dickinson, Penner, Shimasaku,
  Weiner, Kartaltepe, Nakajima, Nayyeri, Stern, Kashikawa, \&
  Spinrad}]{Ono:2012p27651}
Ono, Y., Ouchi, M., Mobasher, B., {et~al.} 2012, The Astrophysical Journal,
  744, 83

\bibitem[{Pentericci {et~al.}(2011)Pentericci, Fontana, Vanzella, Castellano,
  Grazian, Dijkstra, Boutsia, Cristiani, Dickinson, Giallongo, Giavalisco,
  Maiolino, Moorwood, Paris, \& Santini}]{Pentericci:2011p27723}
Pentericci, L., Fontana, A., Vanzella, E., {et~al.} 2011, The Astrophysical
  Journal, 743, 132

\bibitem[{Planck-Collaboration {et~al.}(2013)Planck-Collaboration, Ade,
  Aghanim, Armitage-Caplan, Arnaud, Ashdown, Atrio-Barandela, Aumont,
  Baccigalupi, Banday, Barreiro, Bartlett, Battaner, Benabed, Beno{\^\i}t,
  Benoit-L{\'e}vy, Bernard, Bersanelli, Bielewicz, Bobin, Bock, Bonaldi, Bond,
  Borrill, Bouchet, Bridges, Bucher, Burigana, Butler, Calabrese, Cappellini,
  Cardoso, Catalano, Challinor, Chamballu, Chary, Chen, Chiang, Chiang,
  Christensen, Church, Clements, Colombi, Colombo, Couchot, Coulais, Crill,
  Curto, Cuttaia, Danese, Davies, Davis, de~Bernardis, de~Rosa, de~Zotti,
  Delabrouille, Delouis, D{\'e}sert, Dickinson, Diego, Dolag, Dole, Donzelli,
  Dor{\'e}, Douspis, Dunkley, Dupac, Efstathiou, Elsner, En{\ss}lin, Eriksen,
  Finelli, Forni, Frailis, Fraisse, Franceschi, Gaier, Galeotta, Galli, Ganga,
  Giard, Giardino, Giraud-H{\'e}raud, Gjerl{\o}w, Gonz{\'a}lez-Nuevo,
  G{\'o}rski, Gratton, Gregorio, Gruppuso, Gudmundsson, Haissinski, Hamann,
  Hansen, Hanson, Harrison, Henrot-Versill{\'e}, Hern{\'a}ndez-Monteagudo,
  Herranz, Hildebrandt, Hivon, Hobson, Holmes, Hornstrup, Hou, Hovest,
  Huffenberger, Jaffe, Jaffe, Jewell, Jones, Juvela, Keih{\"a}nen, Keskitalo,
  Kisner, Kneissl, Knoche, Knox, Kunz, Kurki-Suonio, Lagache,
  L{\"a}hteenm{\"a}ki, Lamarre, Lasenby, Lattanzi, Laureijs, Lawrence, Leach,
  Leahy, Leonardi, Le{\'o}n-Tavares, Lesgourgues, Lewis, Liguori, Lilje,
  Linden-V{\o}rnle, L{\'o}pez-Caniego, Lubin, Mac{\'\i}as-P{\'e}rez, Maffei,
  Maino, Mandolesi, Maris, Marshall, Martin, Mart{\'\i}nez-Gonz{\'a}lez, Masi,
  Massardi, Matarrese, Matthai, Mazzotta, Meinhold, Melchiorri, Melin, Mendes,
  Menegoni, Mennella, Migliaccio, Millea, Mitra, Miville-Desch{\^e}nes, Moneti,
  Montier, Morgante, Mortlock, Moss, Munshi, Murphy, Naselsky, Nati, Natoli,
  Netterfield, N{\o}rgaard-Nielsen, Noviello, Novikov, Novikov, O'Dwyer,
  Osborne, Oxborrow, Paci, Pagano, Pajot, Paoletti, Partridge, Pasian,
  Patanchon, Pearson, Pearson, Peiris, Perdereau, Perotto, Perrotta, Pettorino,
  Piacentini, Piat, Pierpaoli, Pietrobon, Plaszczynski, Platania,
  Pointecouteau, Polenta, Ponthieu, Popa, Poutanen, Pratt, Pr{\'e}zeau, Prunet,
  Puget, Rachen, Reach, Rebolo, Reinecke, Remazeilles, Renault, Ricciardi,
  Riller, Ristorcelli, Rocha, Rosset, Roudier, Rowan-Robinson,
  Rubi{\~n}o-Mart{\'\i}n, Rusholme, Sandri, Santos, Savelainen, Savini, Scott,
  Seiffert, Shellard, Spencer, Starck, Stolyarov, Stompor, Sudiwala, Sunyaev,
  Sureau, Sutton, Suur-Uski, Sygnet, Tauber, Tavagnacco, Terenzi, Toffolatti,
  Tomasi, Tristram, Tucci, Tuovinen, T{\"u}rler, Umana, Valenziano, Valiviita,
  Tent, Vielva, Villa, Vittorio, Wade, Wandelt, Wehus, White, White, Wilkinson,
  Yvon, Zacchei, \& Zonca}]{PlanckCollaboration:2013p33616}
Planck-Collaboration, Ade, P. A.~R., Aghanim, N., {et~al.} 2013, eprint arXiv,
  1303, 5076

\bibitem[{Postman {et~al.}(2012)Postman, Coe, Ben{\'\i}tez, Bradley,
  Broadhurst, Donahue, Ford, Graur, Graves, Jouvel, Koekemoer, Lemze,
  Medezinski, Molino, Moustakas, Ogaz, Riess, Rodney, Rosati, Umetsu, Zheng,
  Zitrin, Bartelmann, Bouwens, Czakon, Golwala, Host, Infante, Jha,
  Jimenez-Teja, Kelson, Lahav, Lazkoz, Maoz, McCully, Melchior, Meneghetti,
  Merten, Moustakas, Nonino, Patel, Reg{\"o}s, Sayers, Seitz, \&
  Wel}]{Postman:2012p27556}
Postman, M., Coe, D., Ben{\'\i}tez, N., {et~al.} 2012, The Astrophysical
  Journal Supplement, 199, 25

\bibitem[{Robertson {et~al.}(2013)Robertson, Furlanetto, Schneider, Charlot,
  Ellis, Stark, McLure, Dunlop, Koekemoer, Schenker, Ouchi, Ono, Curtis-Lake,
  Rogers, Bowler, \& Cirasuolo}]{Robertson:2013p27340}
Robertson, B.~E., Furlanetto, S.~R., Schneider, E., {et~al.} 2013, eprint
  arXiv, 1301, 1228

\bibitem[{{Schenker} {et~al.}(2012){Schenker}, {Stark}, {Ellis}, {Robertson},
  {Dunlop}, {McLure}, {Kneib}, \& {Richard}}]{2012ApJ...744..179S}
{Schenker}, M.~A., {Stark}, D.~P., {Ellis}, R.~S., {et~al.} 2012, \apj, 744,
  179

\bibitem[{Schmidt {et~al.}(2014)Schmidt, Treu, Trenti, Bradley, Kelly, Oesch,
  Shull, \& Stiavelli}]{Schmidt:2014p33431}
Schmidt, K.~B., Treu, T., Trenti, M., {et~al.} 2014, submitted to ApJ

\bibitem[{Stark {et~al.}(2011)Stark, Ellis, \& Ouchi}]{Stark:2011p27664}
Stark, D.~P., Ellis, R.~S., \& Ouchi, M. 2011, The Astrophysical Journal
  Letters, 728, L2

\bibitem[{{Taylor} \& {Lidz}(2013)}]{2013MNRAS.tmp.2740T}
{Taylor}, J., \& {Lidz}, A. 2013, \mnras, arXiv:1308.6322

\bibitem[{Trenti {et~al.}(2011)Trenti, Bradley, Stiavelli, Oesch, Treu,
  Bouwens, Shull, MacKenty, Carollo, \& Illingworth}]{Trenti:2011p12656}
Trenti, M., Bradley, L.~D., Stiavelli, M., {et~al.} 2011, The Astrophysical
  Journal Letters, 727, L39

\bibitem[{Treu {et~al.}(2013)Treu, Schmidt, Trenti, Bradley, \&
  Stiavelli}]{Treu:2013p32132}
Treu, T., Schmidt, K.~B., Trenti, M., Bradley, L.~D., \& Stiavelli, M. 2013,
  The Astrophysical Journal Letters, 775, L29

\bibitem[{Treu {et~al.}(2012)Treu, Trenti, Stiavelli, Auger, \&
  Bradley}]{Treu:2012p12658}
Treu, T., Trenti, M., Stiavelli, M., Auger, M.~W., \& Bradley, L.~D. 2012, The
  Astrophysical Journal, 747, 27

\bibitem[{Vanzella {et~al.}(2009)Vanzella, Giavalisco, Dickinson, Cristiani,
  Nonino, Kuntschner, Popesso, Rosati, Renzini, Stern, Cesarsky, Ferguson, \&
  Fosbury}]{Vanzella:2009p29479}
Vanzella, E., Giavalisco, M., Dickinson, M., {et~al.} 2009, The Astrophysical
  Journal, 695, 1163

\bibitem[{Vanzella {et~al.}(2013)Vanzella, Fontana, Zitrin, Coe, Bradley,
  Postman, Grazian, Castellano, Pentericci, Giavalisco, Rosati, Nonino, Smit,
  Balestra, Bouwens, Cristiani, Giallongo, Zheng, Infante, Cusano, \&
  Speziali}]{Vanzella:2013p33637}
Vanzella, E., Fontana, A., Zitrin, A., {et~al.} 2013, eprint arXiv, 1312, 6299

\bibitem[{{Zitrin} {et~al.}(2009){Zitrin}, {Broadhurst}, {Rephaeli}, \&
  {Sadeh}}]{2009ApJ...707L.102Z}
{Zitrin}, A., {Broadhurst}, T., {Rephaeli}, Y., \& {Sadeh}, S. 2009, \apjl,
  707, L102

\end{thebibliography}
\end{document}